\documentclass[%
 reprint,
amsmath,amssymb,
aps,
pre,
]{revtex4-1}

\usepackage{array}
\usepackage{pifont}
\usepackage[usenames, dvipsnames]{color}
\usepackage{tabu}
\usepackage{graphicx}
\usepackage{dcolumn}
\usepackage{bm}

\newcommand{\TE}{\left<TE\right>}

\usepackage{soul,xcolor}				
\setstcolor{red}         				
\usepackage{color,soul}					
\definecolor{red}{rgb}{1,0,0}
\setulcolor{red}

\begin{document}

\preprint{number}

\title{An information-based classification of Elementary Cellular Automata}

\author{Enrico Borriello$^1$}
\email{enrico.borriello@asu.edu}
\author{Sara Imari Walker$^{1,2,3,4}$}
\email{sara.i.walker@asu.edu}
\affiliation{
$^1$Beyond Center for Fundamental Concepts in Science, Arizona State University, Tempe AZ;
$^2$School of Earth and Space Exploration, Arizona State University, Tempe AZ; $^3$ ASU-SFI Center for Biosocial Complex Systems, $^4$ Blue Marble Space Institute of Science
}

\begin{abstract} 
A novel, information-based classification of 
elementary cellular automata
is proposed that circumvents the problems associated with isolating whether complexity is in fact intrinsic to a dynamical rule, or if it arises merely as a product of a complex initial state. Transfer entropy variations processed by the system split the 256 elementary rules into three {\it information classes}, based on sensitivity to initial conditions. These classes form a hierarchy such that coarse-graining transitions observed among elementary cellular automata rules predominately occur within each information-based class, or much more rarely, down the hierarchy.

\end{abstract}

\maketitle

\section{Introduction}

Complexity is easily identified, but difficult to quantify. Attempts to classify dynamical systems in terms of their complexity have therefore so-far relied primarily on qualitative criteria. A good example is the classification of cellular automaton (CA) rules. CA are discrete dynamical systems of great interest in complexity science because they capture two key features of many physical systems: they evolve according to a local uniform rule, and can exhibit rich emergent behavior even from very simple rules \cite{Israeli2006}. Studying the dynamics of CA therefore can provide insights into the harder problem of how it is that the natural world appears so complex given that the known laws of physics are local and (relatively) simple. However, in the space of CA rules, interesting emergent behavior is the exception rather than the rule. This has generated wide-spread interest in understanding how to segregate those local rules that generate rich, emergent behavior -- including coherent structures such as gliders and particles or even computational universality \cite{general_refs_1,general_refs_2,general_refs_3} -- from those that do not.
A complication arises in that the complexity of the output of a CA rule is often highly dependent on that of the input state, making it difficult to disentangle emergent behavior that is a product of the initial state from that which is {\it intrinsic} to the rule. This has resulted in ambiguity in classifying the intrinsic complexity of CA rules as one must inevitably execute a CA rule with a particular initial state in order to express its complexity (or lack thereof).

\begin{figure*}[!t]
\begin{minipage}[t]{0.3\textwidth} 
(a) single cell input \\
\includegraphics[width=\columnwidth]{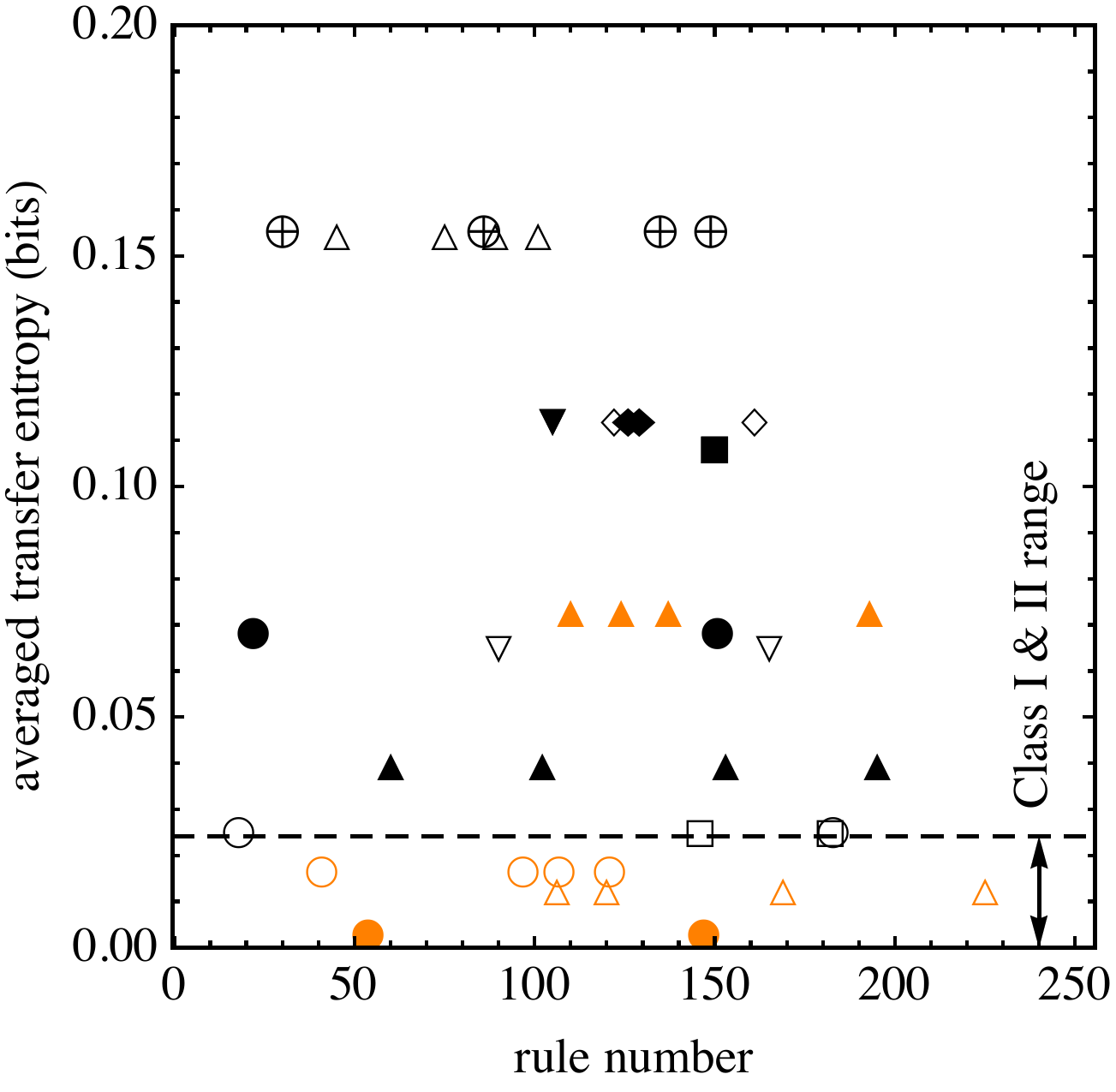} 
\end{minipage} \quad
\begin{minipage}[t]{0.3\textwidth}
(b) random input \\
\includegraphics[width=\columnwidth]{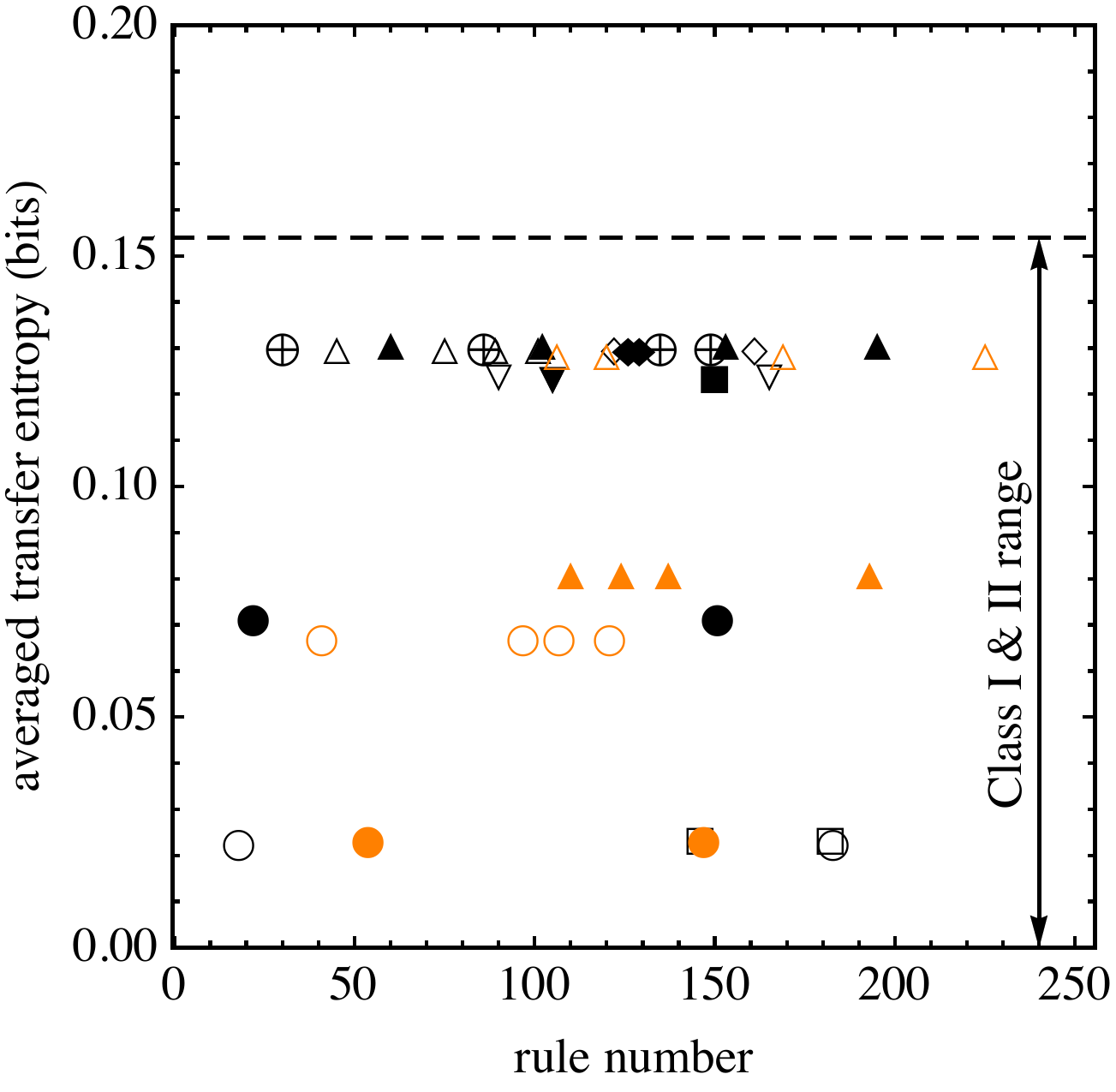} 
\end{minipage} \quad
\begin{minipage}[t]{0.3\textwidth}
legend\\[4mm]
\includegraphics[width=\columnwidth]{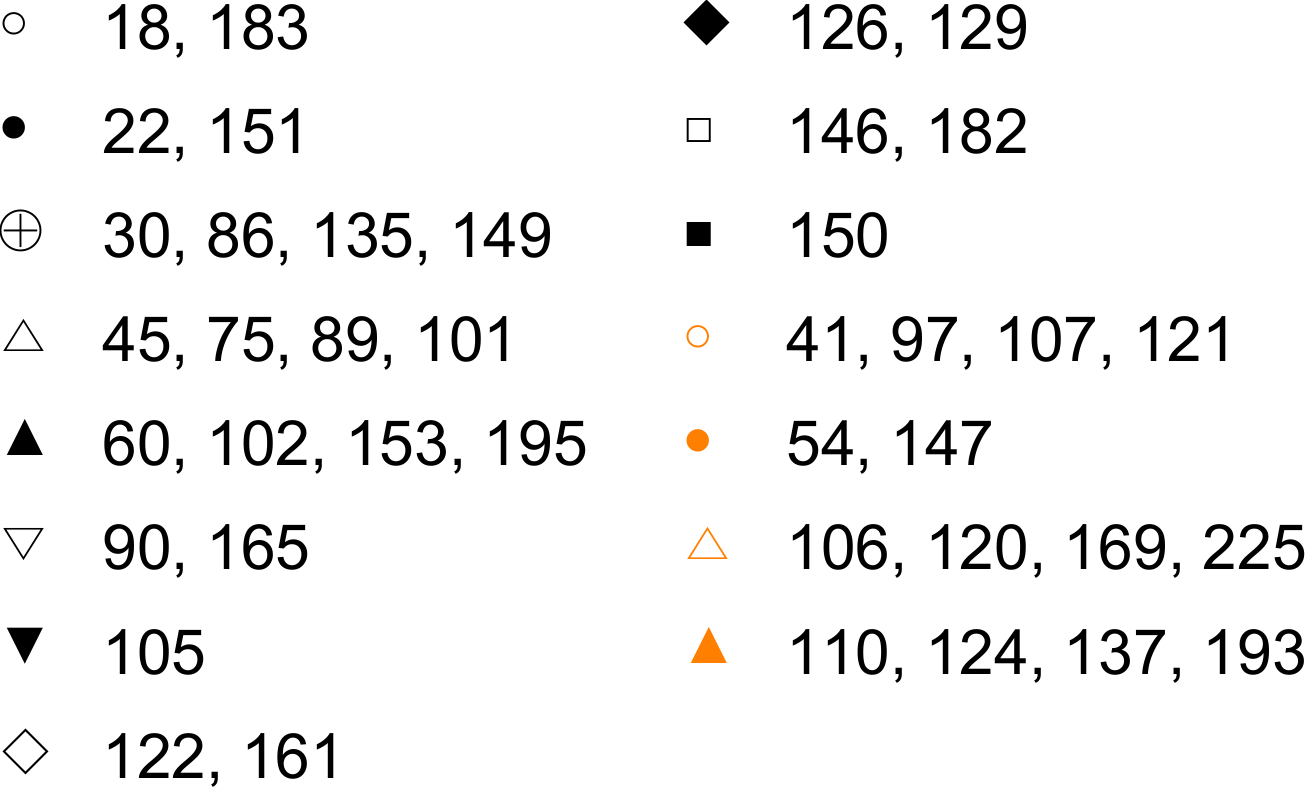}
\end{minipage}
\caption{$\TE$ for each rule, shown for two different inputs: Single cell (a), and random (b). Equivalent rules are represented by a common marker. The dashed line corresponds to the highest value of $\TE$ for any Wolfram Class I or II rule. Individual markers are shown only for Wolfram Class III (black) and IV (orange) rules.}
\label{single-cell-input}
\end{figure*}

One of the first attempts to classify CA rules was provided by Wolfram in his classification of elementary cellular automata (ECA) \cite{Wolfram1984}. ECA are some of the simplest CA and are 1-dimensional with nearest-neighbor update rules operating on the two-bit alphabet $\{\blacksquare, \square\}$. Despite their simplicity, ECA are known to be capable of complex emergent behavior. Initializing an ECA in a random state leads some rules to converge on fixed point or oscillatory attractors, while others lead to chaotic patterns that are computationally irreducible, such that their dynamics are impossible to predict from the initial state and chosen rule without actually executing the full simulation. Based on these diverse behaviors, Wolfram identified four distinct complexity classes.  Class I CA are those whose evolution eventually leads to cells of only one kind. Class II CA lead to scattered stable or oscillating behaviors. Class III CA show an irreducibly chaotic pattern. Class IV CA can exhibit any of the previous behaviors simultaneously, and seem to posses the kind of complexity that lies at the interface between mathematical modeling and life studies \cite{VonNeumann1966}. Wolfram's classification stands as a milestone in the understanding of CA properties, and still represents the most widely adopted classification scheme. Nonetheless, its qualitative nature represents its main limitation, as a result of automata being able to show characteristics typical of different classes for different initial states \cite{Baldwin98,Culik1988}. 

In this paper we report on experiments demonstrating quantifiable classification of the intrinsic complexity of ECAs, by exploiting the main weakness that has plagued earlier attempts at their quantitative classification. That is, we explicitly utilize the sensitivity of the expressed complexity of ECA rules to the initial input state. Our approach requires a quantitative measure of `complexity'. In recent years there has been increasing interest in using information-theoretic tools to quantify the complexity of dynamical systems, particularly in the context of understanding biological systems \cite{Bialek, WKD2016}, which represent the most complex physical systems known. In this context, one widely utilized measure is transfer entropy ($TE$), Schreiber's measure of the directed exchange of information between two interacting parts of a dynamical system \cite{Schreiber2000,Lizier}. In what follows, we adopt TE as a candidate quantitative selection criterion to classify the intrinsic complexity of ECAs. 

The paper is structured as follows. In section II we start from the simplest, non-trivial, initial configuration of an ECA, and use it to identify the dynamical rules able to produce a complex output by virtue of their intrinsic complexity. In section II we repeat our analysis for more general inputs, and identify those outputs whose complexity is instead inherited by the complexity of the input, as opposed to the rule. We then classify  ECA rules according to the maximum degree of variability of the output they produce for varying inputs. As we will see, three quantitatively and qualitatively distinct classes will emerge. In section III we show that this classification induces a partially ordered hierarchy among the rules, such that coarse-graining a ECA of a given class yields an ECA of the same class, or simpler \cite{Israeli2006}. We conclude by proposing further applications of the classification method presented.

\section{Intrinsic complexity}

We first generated time series for the simplest initial 
state, and for each of the $256$ possible ECA rules, which we numerically label following Wolfram's heuristic numbering scheme \cite{Wolfram1983,Wolfram2002}.
Our initial state has of the form
\[
  \underbrace{\square \dots \square}_{50-\text{times}}
  \blacksquare
  \underbrace{\square \dots \square}_{50-\text{times}}
\]
or the equivalent one obtained through a $\square\leftrightarrow \blacksquare$ conjugation. For example, rules 18 and 183 are equivalent under a $\square\leftrightarrow \blacksquare$ conjugation. The input $ \square \dots \square \blacksquare \square \dots \square $ is updated using rule 18, and the conjugated input $\blacksquare \dots \blacksquare \square \blacksquare \dots \blacksquare $ using rule 183. The association of input $\rightarrow$ rule is chosen heuristically as the one that maximizes the averaged TE over a space-time patch, as described in the next paragraph. Here and in the following, periodic boundary conditions are enforced. Therefore, the specific location of the different color cell in the the input array is irrelevant.

We evaluate TE over a region of the CA that is in causal contact with the initial input at each point in space and time. To do so, we evolve the CA for 250 time-steps and then remove the first 50 time steps from each generated time series. This ensures that the observed dynamics over the relevant space-time patch are driven by both the chosen rule {\it and} the initial state. The resulting time series is then used to evaluate the $101^2$ values $TE_{y \rightarrow x}$. 
For each rule, the average value of $\{TE_{y \rightarrow x}\}$ (averaged over $x$ and $y$), denoted $\left<TE\right>$, is shown Fig. \ref{single-cell-input} (a). Equivalent rules, like rule 30, 86, 135, and 149 produce the same values of $\left<TE\right>$, and are represented by a common marker.
For clarity, individual Wolfram Class I and II rules are not shown, and are instead replaced by a dashed line corresponding to the highest $\left<TE\right>$ of any individual Class I or II rule. All Class III rules lie above the range of Class I and II. Interestingly, with only the exception of rule 110 and its equivalent rules, all Class IV rules lie within the range of Class I and II rules. 

The single cell input considered is extremely rare within the space of all possible initial inputs. The number of black cells in a state randomly extracted among the $2^{101}$ different possible inputs follows a binomial distribution, meaning that states containing 50 or 51 black/white cells are about $2\times10^{27}$ times more likely than our initially selected input. Our motivation for considering the single cell input first is that it automatically excludes many trivial rules from our search for the complex ones. 
Rules that duplicate the input in a trivial way, or annihilate it to produce all $\blacksquare$ or all $\square$, naturally yield $\left<TE\right>\simeq0$. 
This is the same approach that has been recently assumed in algorithmic complexity based classification of CAs \cite{Zenil2013}. It has the advantage of selecting many rules according to their intrinsic complexity, and not the one carried by the input. 

\section{Inherited Complexity and Information-Based Classification}

A drawback of choosing the single-cell input of section I is that many Class IV rules now look simpler than they truly are. Class IV rule 106 is a good example. For the simple input of a single black bit, rule 106 functions to shift the black cell one bit to the left at each time-step, generating a trivial trajectory. However, in cases where the input allows two black cells to interact, the full potential of rule 106 can be expressed. This is an explicit example of the sensitivity of the behavior of some ECA rules, which inherit the complexity of their input, as discussed in the introduction. 

Let us therefore next consider a more generic input, randomly selected among the $2^{101}\simeq 2.5\times10^{30}$ different possibilities. As the input is no longer symmetric, we now need to consider reflections in selecting the equivalent input $\rightarrow$ rule associations. The scenario changes completely in this case, as shown in Fig. \ref{single-cell-input} (b), where the highest values of $\left<TE\right>$ correspond to Class II rules, including 15, 85, 170, and 240, which generate trivial, wave-like behaviors. These rules behave like rule 106 (especially 170 and 240) when initialized with a single cell input, but they do not contribute any new, emergent non-trivial features when nearby sites interact. For all purposes they appear {\it less} intrinsically complex.  
With only the exception of rule 110 and its equivalent rules, Class IV rules behave like many rules of Class II, and exhibit a large increase in complexity, as qualitatively observed and also as captured by $\left<TE\right>$, in response to a more complex input. 

\begin{figure}[!t]
\centering
\includegraphics[width=.85\columnwidth]{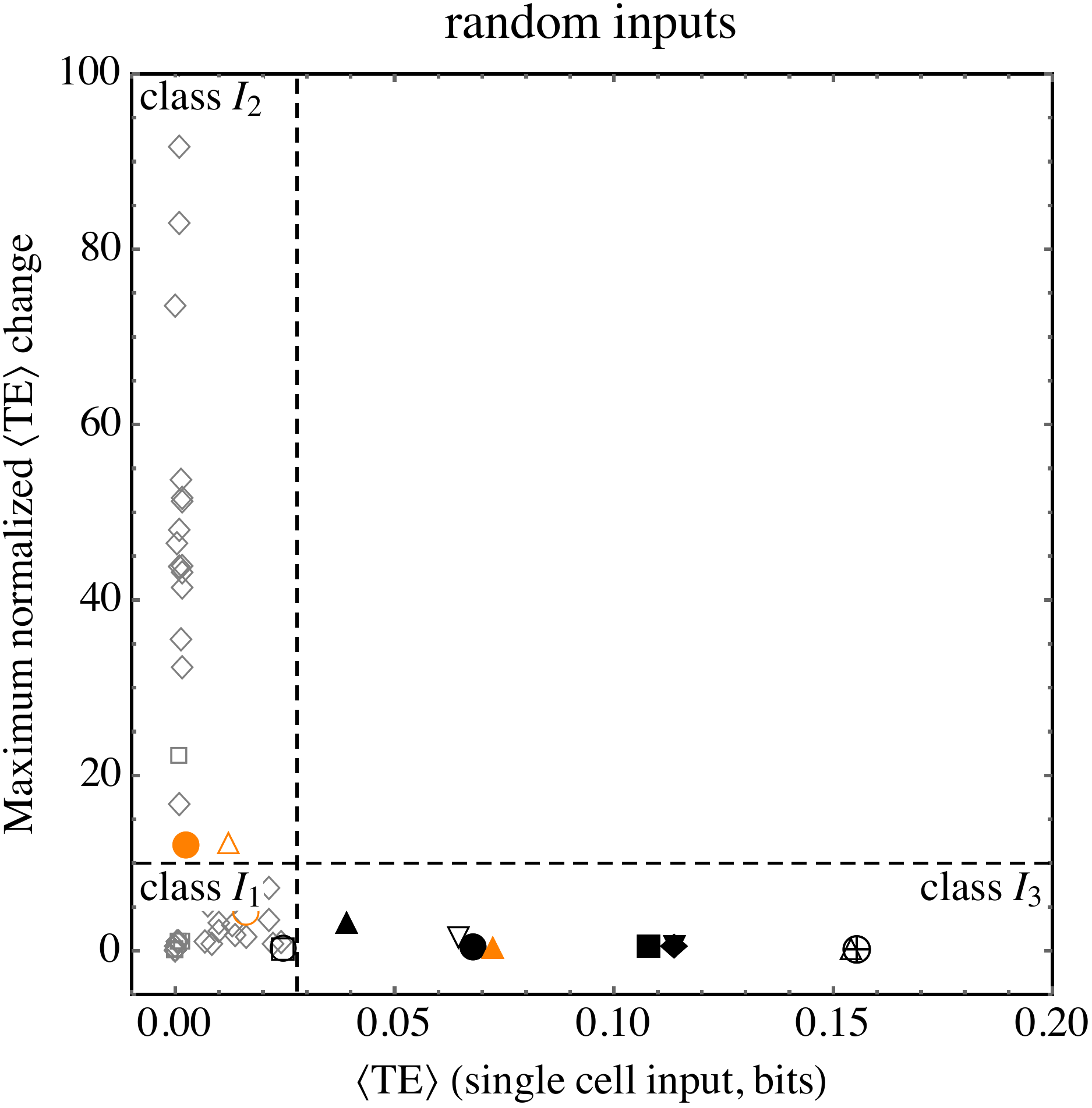}
\caption{Maximum normalized change in $\TE$ as a function of {$\TE_{\textrm{1}}$}. Equivalent rules are represented by a single marker, colored as in Fig. \ref{single-cell-input} but with Wolfram Classes I and II in gray. Rules above the horizontal dotted line can undergo a change of at least one order of magnitude in $\TE$. The vertical dotted line is the same as in Fig. \ref{single-cell-input} (a).}
\label{percent-change}
\end{figure}

It is worth noting that all of the rules whose initial value of {$\TE$} lay above the upper limit for Classes I and II of $\left<TE\right> =0.024$ bits in the simplest input scenario, still have calculated $\left<TE\right>$ above this value under a change of input. In particular, the rules with the highest values of {$\TE$} are not significantly affected by the change in the input. Let us naively use the upper limit for Class I and II rules emerging from Fig. \ref{single-cell-input} (a) as the border line between what we call {\it low} and {\it high} values of {$\TE$}. We can summarize the changes under varying the input as follows. There are rules whose value of {$\TE$} is low for both the inputs. There are rules whose value of {$\TE$} is low in the simplest case, but high in the more complex one. And, finally there are also rules with a high value of {$\TE$} in both cases. The interesting point is that {\it we find no rule whose value of {$\TE$} is high for the simplest input, and low for the random input}. That is, there exist no rules that generate complexity for simple inputs, but annihilate it for complex ones. The relevance of this observation lies in the fact that it enables classification in terms of the shift of $\left<TE\right>$ over a space-time region in response to a change in its input. This is to say that this feature takes advantage of the main limitation that makes quantitative classifications of ECAs so difficult, {\it i.e.} ECA sensitivity to their initial state, and exploits it in order to achieve such classification.

To confirm this is indeed viable, we must consider more than just two inputs. We therefore first randomly selected twenty different ones. Being interested in how much {$\TE$} can vary as we vary the input, for each rule we selected the maximum absolute value of the change of {$\TE$} between the random inputs ($\TE_{\textrm{r}}$) and the single cell input ($ \TE_{\textrm{1}}$) considered before, normalized to the value for the single cell input:
\[
\textrm{max}\left\{
\frac{  \left|\TE_{\textrm{r}} - \TE_{\textrm{1}}\right|  }{ \TE_{\textrm{1}} }
\right\}_
{\begin{scriptsize}\begin{array}{l}
\textrm{random}\\
\textrm{inputs}
\end{array}\end{scriptsize}} \ .
\]

The results are shown in Fig. \ref{percent-change}, where the maximum normalized change is plotted as a function of {$\TE_{\textrm{1}}$}. Equivalent rules are represented by a single marker. Rules above the horizontal dotted line can undergo a change in $\TE$ of at least one order of magnitude. The vertical dotted line denotes the highest value of {$\TE_{\textrm{1}}$} for Wolfram Classes I and II, exactly as in Fig. \ref{single-cell-input} (a). The region to the right of the vertical line and above the horizontal line, is void of any ECA rules. Points in that region would correspond to values of {$\TE$} that are both high for the simplest input, and capable of high variation, {\it e.g.} CA rules that can annihilate the complexity of the input, which we do not observe.
This feature yields a distinctive L-shape in the distribution of rules with the rules in each region sharing well defined properties.

As a result, an information-based classification of ECAs can be given as follows:\\

\noindent 
{\it \bf Class} I$_1$: $\left<TE\right>$ is zero, or very small, for the simplest input, and stays so for complex inputs. This is the most populated class, including almost all Wolfram Class I and II rules, rule 18, 146 and their equivalent Class III rules, as well as rule 41 and its equivalent Class IV rules.\\

\noindent 
{\it \bf Class} I$_2$: $\left<TE\right>$ is small for the simplest input, but it experiences a drastic change (one order of magnitude or more) when the input is complex. This is the case for many  Wolfram Class II and some Class IV rules ({\it e.g.} 54, 106 and their equivalent rules).\\

\noindent 
{\it \bf Class} I$_3$: $\left<TE\right>$ has a high value for the simplest input, and this value is approximately unaffected by a change in the input. Most Wolfram Class III rules belong to this class, as well as Class IV rule 110 and its equivalent rules.\\

Randomly sampling inputs leads to a bias favoring nearly fifty/fifty distributions of black/white cells, due to their binomial distribution. We therefore also verified this classification using a different distribution of inputs, where the number of black cells is increased in a regular way from 2 to 50, while the specific positions of these cells in the input array are random. We considered 20 different inputs, each containing exactly 2, 5, 7, 10, 12, $\dots$ 47, and 50 black cells (higher numbers are not considered due to the $\square\leftrightarrow \blacksquare$ conjugation). Apart from minor shifts of the data points, applying the same procedure as above yields exactly the same classification as Fig. \ref{percent-change}, indicating that our classification scheme is robust. 

\begin{table}[!b]
\bgroup
\def\arraystretch{1.2}
\begin{tabular}{|rrrrrrrrrr|}
\multicolumn{10}{l}{\bf Class I$_1$}\\ \hline
0 & 1& 3& 4& 5&  8& 9& 12& 13 & {\bf 18}\\
 19& 23& 25& 26& 28& 29& 32& 33& 35 & 36\\
 37& 40& {\bf\color{Orange}41}& 44& 50& 51& 57& 58& 62 & 72\\
 73& 76& 77& 78& 94& 104& 108& 128& 130 & 132\\
 134& 136& 140& {\bf 146}& 152&   154& 156& 160& 162 & 164\\
 178 & 200 & 204 & 232 & & & & & &\\ \hline
\multicolumn{10}{l}{\vspace{-3mm}}\\  
\multicolumn{10}{l}{\bf Class I$_2$}\\ \hline 
2 & 6& 7&10& 11& 14& 15& 24& 27& 34\\
38& 42& 43& 46& {\bf\color{Orange}54}& 56& 74& {\bf\color{Orange}106}&138& 142\\
 168 & 170 & 172 & 184 & & & & & & \\ \hline
\multicolumn{10}{l}{\vspace{-3mm}}\\ 
\multicolumn{10}{l}{\bf Class I$_3$}\\ \hline
{\bf 22} & {\bf 30} & {\bf 45} & {\bf 60} & {\bf 90} & {\bf 105} & {\bf\color{Orange}110} & {\bf 122} & {\bf 126} & {\bf 150} \\ \hline
\end{tabular}
\egroup
\caption{Information based classification of ECA, with only the 88 rule equivalency classes shown (represented by the lowest number rule). Wolfram Class III rules are denoted in black, boldface type, and Class IV in orange.}
\label{Iclasses}
\end{table}

\begin{figure}
\centering
\includegraphics[width=\columnwidth]{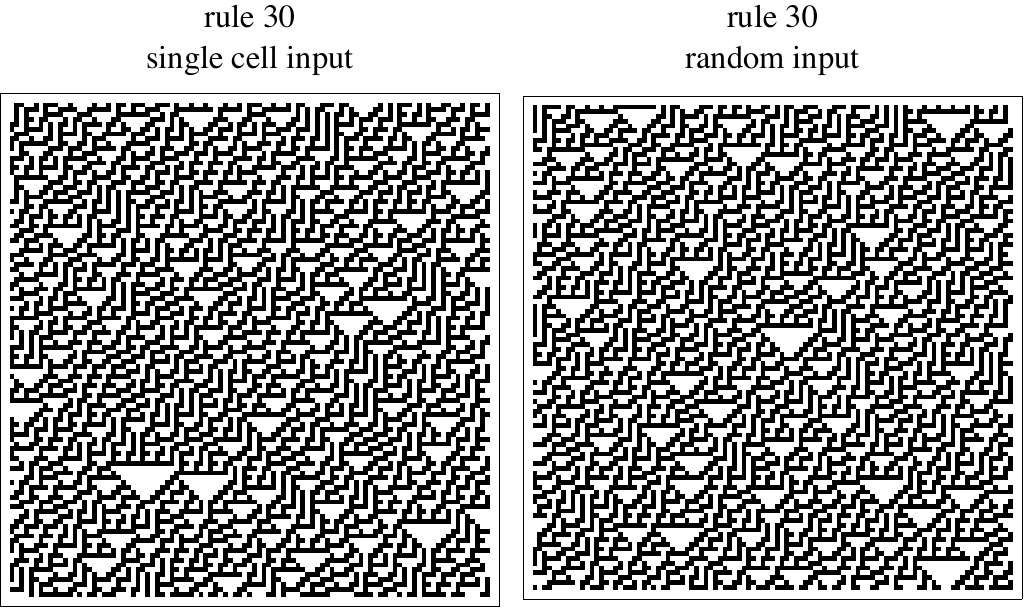}
\caption{Time series for rule 30 generated by the two initial states of Fig. \ref{single-cell-input}. The similarity of the two outputs, each yielding color domains with a typical linear size of about $\sim 10$ blocks, does not depend on the size of the CA and is responsible for the stability of class I$_3$ rules to changes in input.}
\label{grid}
\end{figure}

We stress the importance of considering a {\it large} system (order $\sim100$ cells) as opposed to a much smaller one ({\it e.g.} order $\sim10$). While for the latter a scan over the entire space of inputs is computationally feasible (and indeed we performed these experiments), it hides one of the main features enabling information-based classification -- the existence of class I$_3$ rules, which form the most stable class with respect to our TE based complexity measure. Class I$_3$ rules produce time series largely independent of the initial state --exemplified by rule 30 in Fig. \ref{grid}-- a feature evident only in larger systems. 

\section{Coarse-Graining and the Information Hierarchy}

Perhaps the most interesting feature of our quantitative classification is that Wolfram Class III and Class IV rules are distributed over different information-based classes. This behavior looks less surprising in the light of the coarse-graning transitions among ECA uncovered by Israeli and Goldenfeld \cite{Israeli2006}. One important aspect of the physical world is that coarse-grained descriptions often suffice to make predictions about relevant behavior.  Noting this, Israeli and Goldenfeld adopted a systematic procedure and successfully coarse-grained 240 of the 256 ECA rules, many to other ECA rules. Importantly, the rule complexity was never observed to increase under coarse-graining, introducing a partial ordering among CA rules. 

\begin{figure*}[t]
\begin{minipage}{.65\textwidth}
\centering
\includegraphics[width=\textwidth]{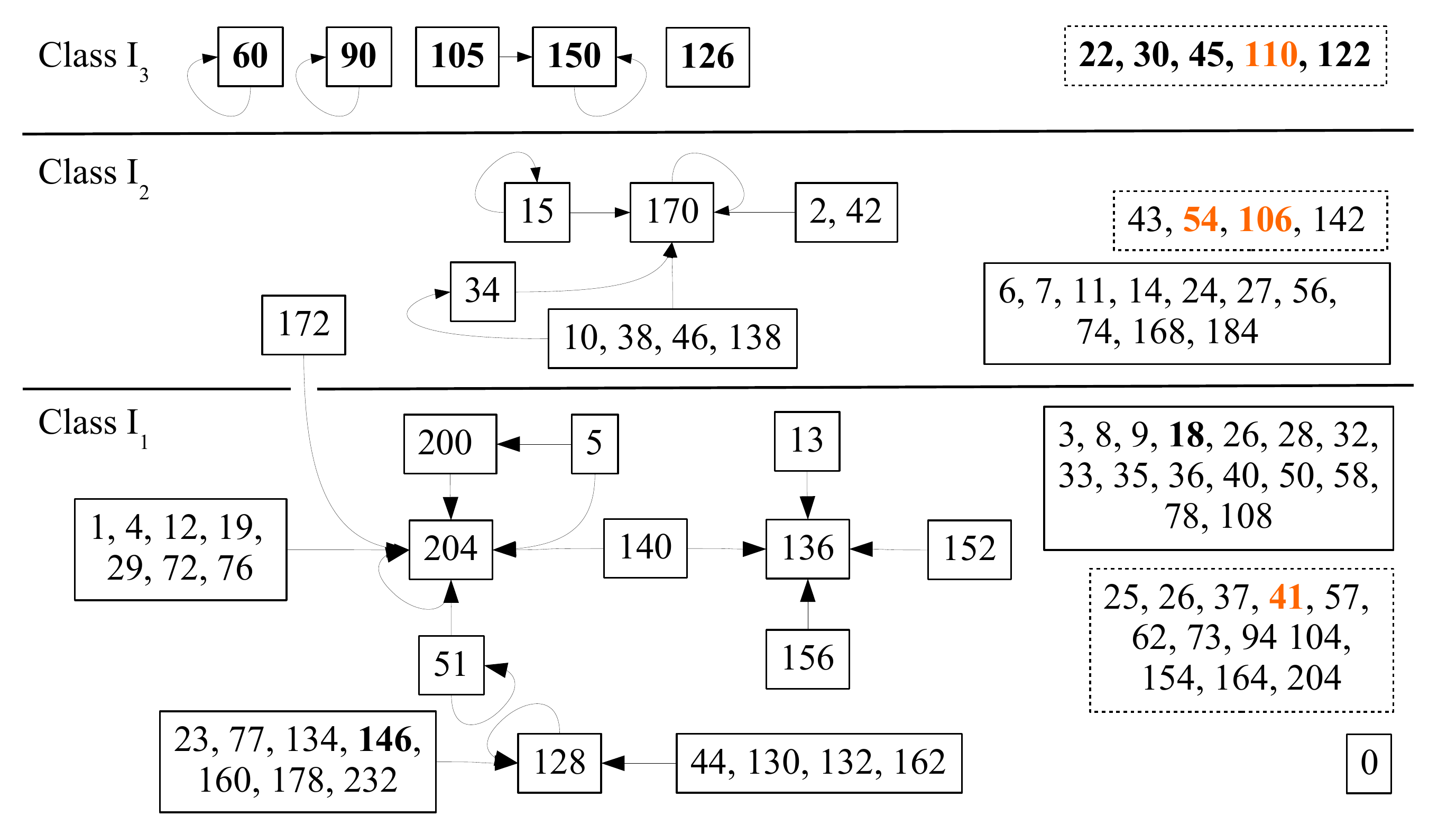}
\end{minipage}
\begin{minipage}[t]{.3\textwidth}
\caption{Coarse-graining transitions within ECA found in \cite{Israeli2006}, mapped to a hierarchy of the information-based classes identified here. Only the 88 rule equivalency classes are shown (represented by the lowest number rule). For simplicity, transitions to rule zero are not shown.}
\label{coarse-graining}
\end{minipage}
\end{figure*}

The same ordering emerges from our information-based classification, as shown in Fig. \ref{coarse-graining}, where arrows indicate coarse-graining transitions uncovered in \cite{Israeli2006}. These transitions introduce a fully ordered hierarchy $I_3 \rightarrow I_2 \rightarrow I_1$ such that coarse-graining is never observed to move up the hierarchy, and the vast majority of rules may only undergo coarse-graining transitions within the {\it same information class}. An example is Wolfram Class III rule 146, which is in $I_1$ because it can be non-trivially coarse-grained to $I_1$ rule 128 (a Wolfram Class I rule), due to a shared sensitivity of the information-processed in a given space-time patch to its input state. We can therefore conclude that the coarse-graining hierarchy is defined by conserved informational properties: more complex rules by Wolfram's classification can appear lower in the hierarchy if they can be coarse-grained to less complex ones with common sensitivity to the input state. 

\section{Conclusions}

Physical systems evolve in time according to local, uniform rules, but can nonetheless exhibit complex, emergent behavior. Complexity arises either as a result of the initial state or rule, or some combination of the two. This ambiguity has confounded previous efforts to isolate complexity {\it intrinsic} to a given rule. In this work we introduced a quantitative, information-based, classification of ECA. The classification scheme proposed circumvents the difficulties arising due to the sensitivity of the expressed complexity of ECA rules to their initial state.
The (averaged) directed exchange of information (TE) between the individual parts of an ECA is used as a measure of its complexity. The identification of the single-cell input (section II) as the non-trivial state with the least complexity was assumed as a working hypothesis, and provides a reference point for our analysis of the degree of variability of the complexity of ECA rules for varying inputs. We identified three distinct class based on our analysis, which vary in their sensitivity to initial conditions. Class $I_1$ ECA always process little information, Class $I_3$ always process high information, and Class $I_2$ can be low or high depending on the input.  It is only for class $I_3$ that the expressed complexity is intrinsic, and not a product of the complexity carried by the input. 

The most complex rules by our analysis are in class $I_3$, which includes the majority of Wolfram's class III rules, and class IV rule 110 and its equivalent rules. These rules form a closed group under the coarse-graining transitions found in \cite{Israeli2006}. The truly complex rules are those that remain complex even at a macro level of description, with behavior that is not sensitive to initial state. Lack of sensitivity to initial conditions is a feature of the robustness of biological systems. 
It is interesting that this is the most complex case here, and that the robustness arises as a result of preserving the information-processed in a particular space-time patch, independent of the initial state. 
\section*{Acknowledgments} 
The authors thank Douglas Moore and Hector Zenil for their helpful comments. This work was supported by a grant from the Templeton World Charity Foundation.


\end{document}